\documentclass[twocolumn,showpacs,preprintnumbers,amsmath,amssymb]{revtex4}

\usepackage{graphicx}

\begin{document}

\title{The $s=\frac{1}{2}$ Antiferromagnetic Heisenberg Model on Fullerene-Type
Symmetry Clusters}

\author{N. P. Konstantinidis$^{*}$}
\affiliation{Department of Physics and Department of Mathematics, University of
             Dublin, Trinity College, Dublin 2, Ireland}

\date{\today}

\begin{abstract}
The $s_{i}=\frac{1}{2}$ nearest neighbor antiferromagnetic Heisenberg model is
considered for spins sitting on the vertices of clusters with the connectivity
of fullerene molecules and a number of sites $n$ ranging from $24$ to $32$.
Using the permutational and spin inversion symmetries of the Hamiltonian the
low energy spectrum is calculated for all the irreducible representations of
the symmetry group of each cluster. Frustration and connectivity result in
non-trivial low energy properties, with the lowest excited states being
singlets except for $n=28$. Same hexagon and same pentagon correlations are the
most effective in the minimization of the energy, with the $n=32-D_{3h}$
symmetry cluster having an unusually strong singlet intra-pentagon correlation.
The magnetization in a field shows no discontinuities unlike the icosahedral
$I_h$ fullerene clusters, but only plateaux with the most pronounced for
$n=28$. The
spatial symmetry as well as the connectivity of the clusters appear to be
important for the determination of their magnetic properties.
\end{abstract}

\pacs{75.10.Jm Quantized Spin Models, 75.50.Ee Antiferromagnetics,
      75.50.Xx Molecular Magnets}

\maketitle

\section{Introduction}
\label{sec:1}

The antiferromagnetic Heisenberg model (AHM) has been the object of intense
investigation for some time now as a prototype of strongly correlated
electronic behavior. The effects of frustration, quantum fluctuations and low
dimensionality can lead to new phases differing from conventional order and
possessing a non-trivial low energy spectrum
\cite{Misguich03,Lhuillier02,Lhuillier01}. Small magnetic clusters
provide an excellent testing ground for the validity of the AHM as well as
other theoretical models, as oftentimes its low energy properties can be
computed on these structures and its predictions can be directly tested against
experiments \cite{Schnack04}.

The fullerenes are a class of three-fold coordinated molecules consisting of
$\frac{n}{2}-10$ hexagons and $12$ pentagons, with $n$ their number of
vertices (or sites) \cite{Fowler95}. With increasing $n$ their shape
resembles more and more the honeycomb lattice, albeit in closed form, with the
pentagons playing the role of structural impurities. Frustration decreases
with $n$, as the unfrustrated
hexagons dominate in number the frustrated pentagons, while the distribution
of the latter determines the symmetry group of the molecule.
The properties of the AHM have been computed for the smallest element of the
family, the dodecahedron, which consists only of pentagons and
belongs to the icosahedral point symmetry group $I_h$.
For classical spins the signature of
frustration is very strong, generating three magnetization discontinuitites in
an external magnetic field, unexpectedly for a model lacking magnetic
anisotropy \cite{Coffey92,NPK07}. In the full quantum limit
where the individual spin magnitude $s_i=\frac{1}{2}$, the low energy spectrum
consists of singlets, absent in unfrustrated systems \cite{NPK05-1}. More
unconventional behavior is displayed by the
magnetization which is discontinuous in
an external field as in the classical case, and the
specific heat which has a two-peak structure as a function of temperature. For
$s_i=1$ non-magnetic excitations are still present inside the singlet-triplet
gap, and now there are two magnetization discontinuities in a field.
Similar behavior with magnetization discontinuities was also
found for the AHM on larger fullerene molecules of
$I_h$ symmetry for classical and $s_i=\frac{1}{2}$ spins \cite{NPK07}. There
is strong evidence that this behavior is persistent for $I_h$
symmetry in the
$n \to \infty$ limit and survives asymptotically close to the zero and
saturation fields, even though the number of hexagons strongly dominates the 12
frustrated pentagons.

\begin{figure}
\includegraphics[width=1.65in,height=2in]{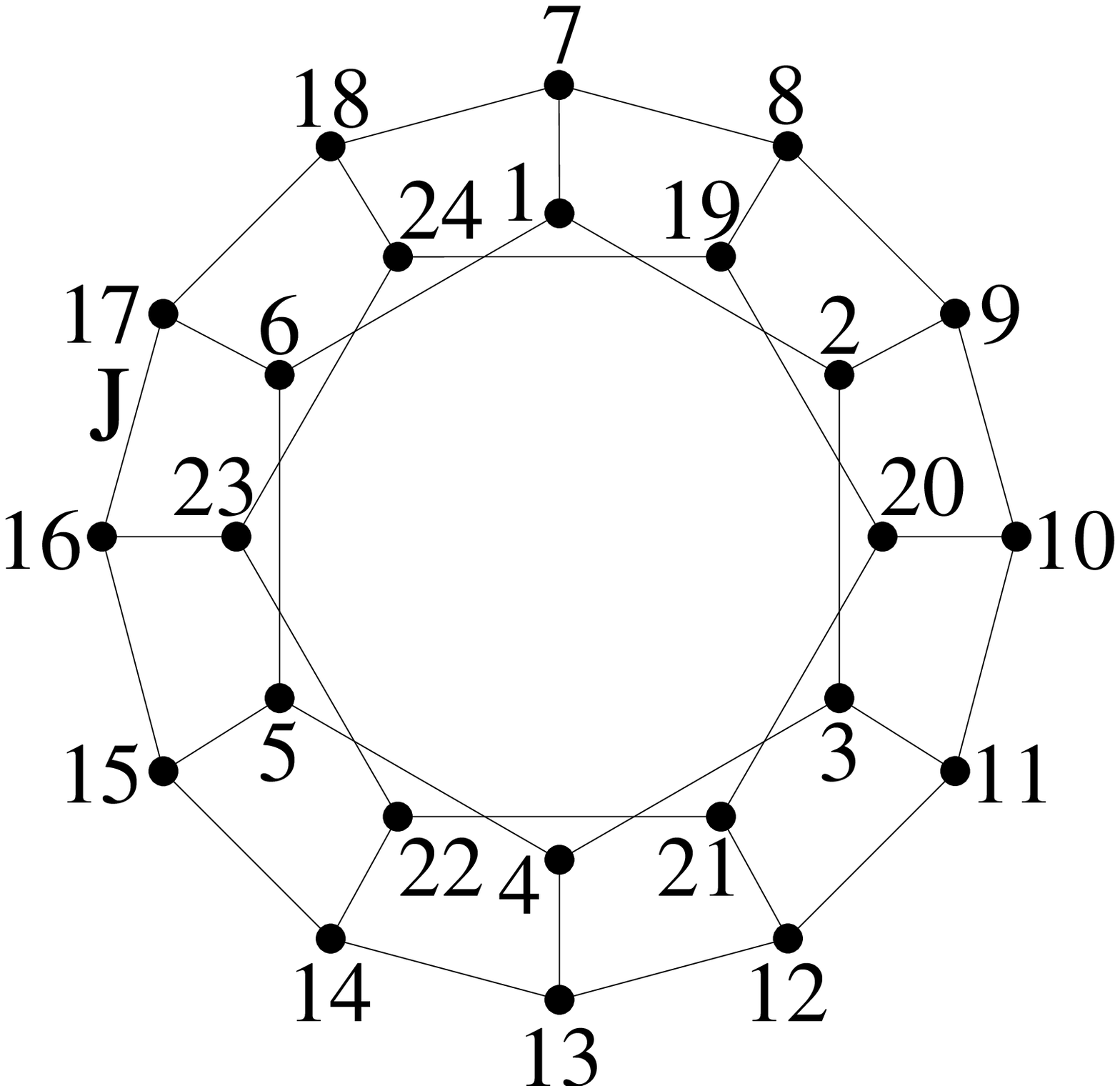}
\includegraphics[width=1.65in,height=2in]{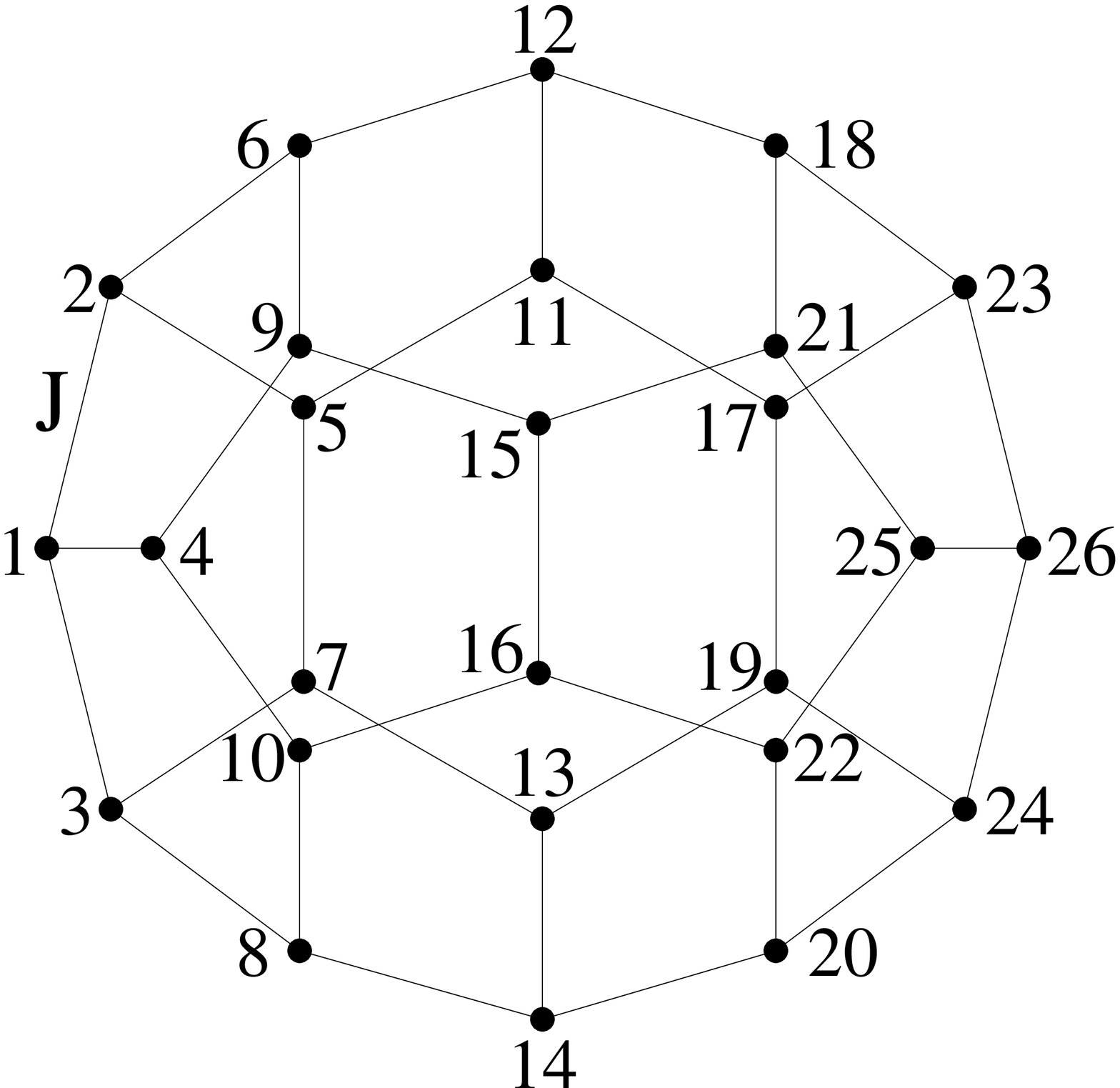}
\text{(a) \hspace{103pt} (b)}

\vspace{10pt}
\includegraphics[width=1.65in,height=2in]{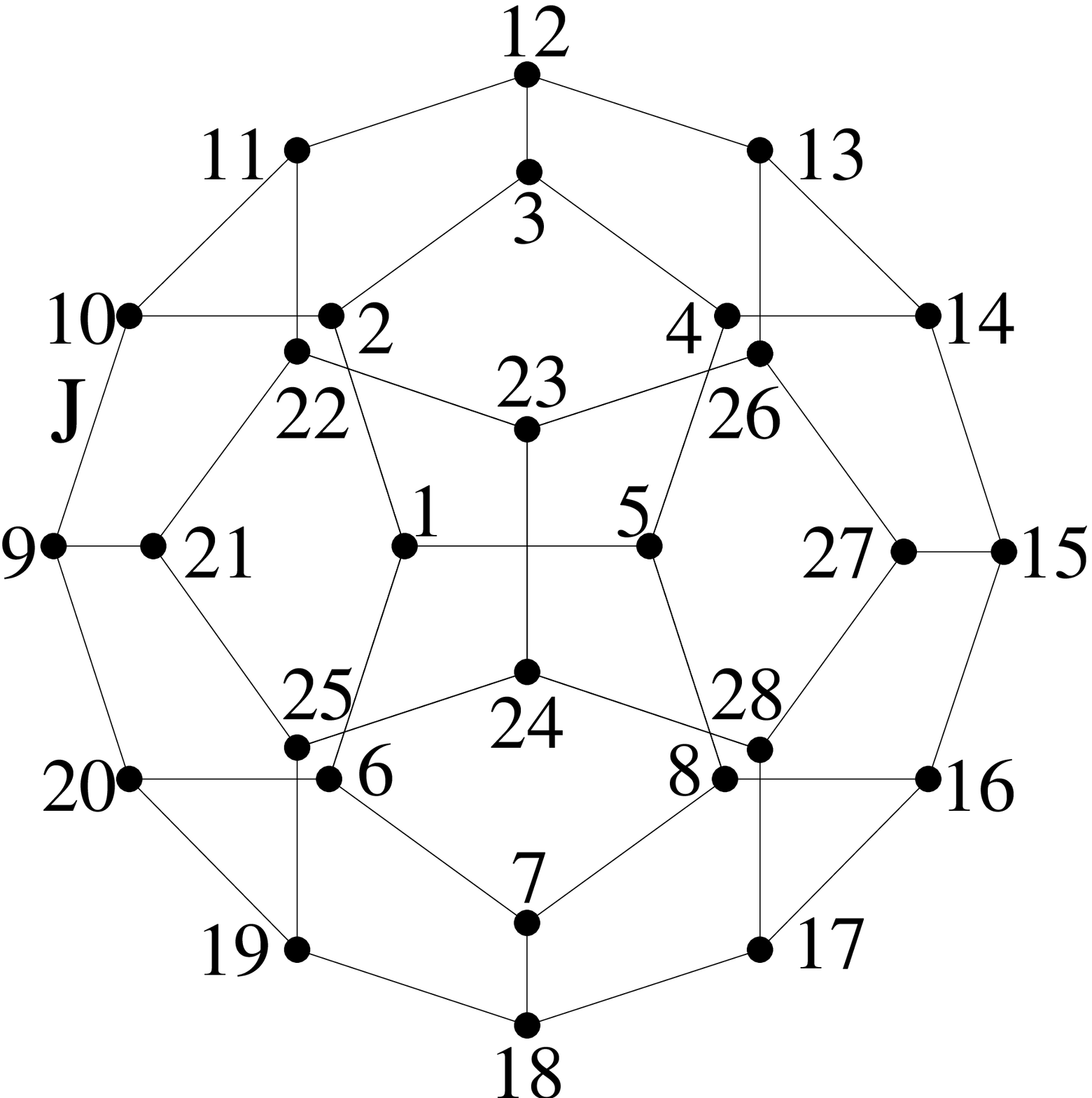}
\includegraphics[width=1.65in,height=2in]{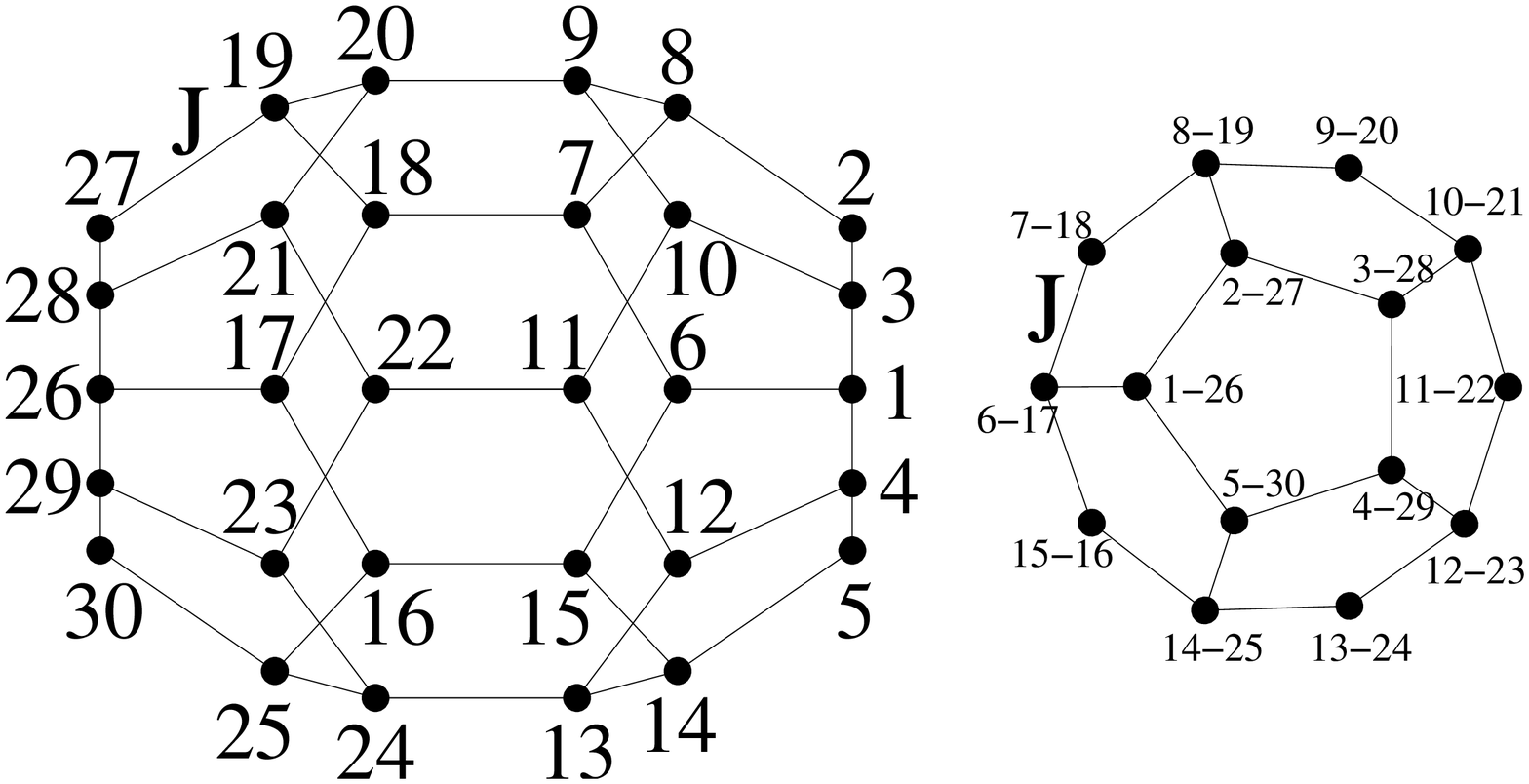}
\text{(c) \hspace{103pt} (d)}

\vspace{10pt}
\includegraphics[width=1.65in,height=2in]{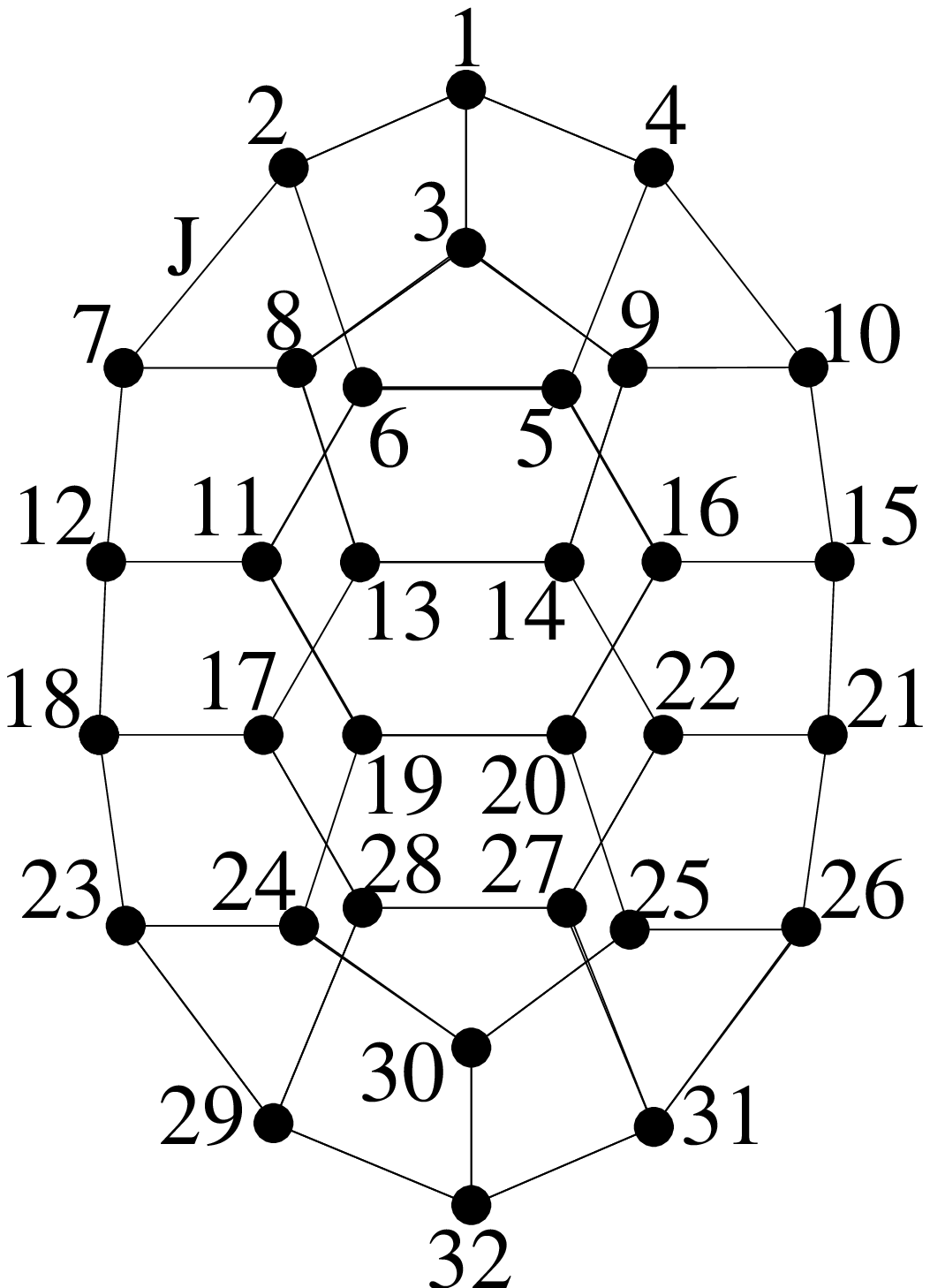}
\includegraphics[width=1.65in,height=2in]{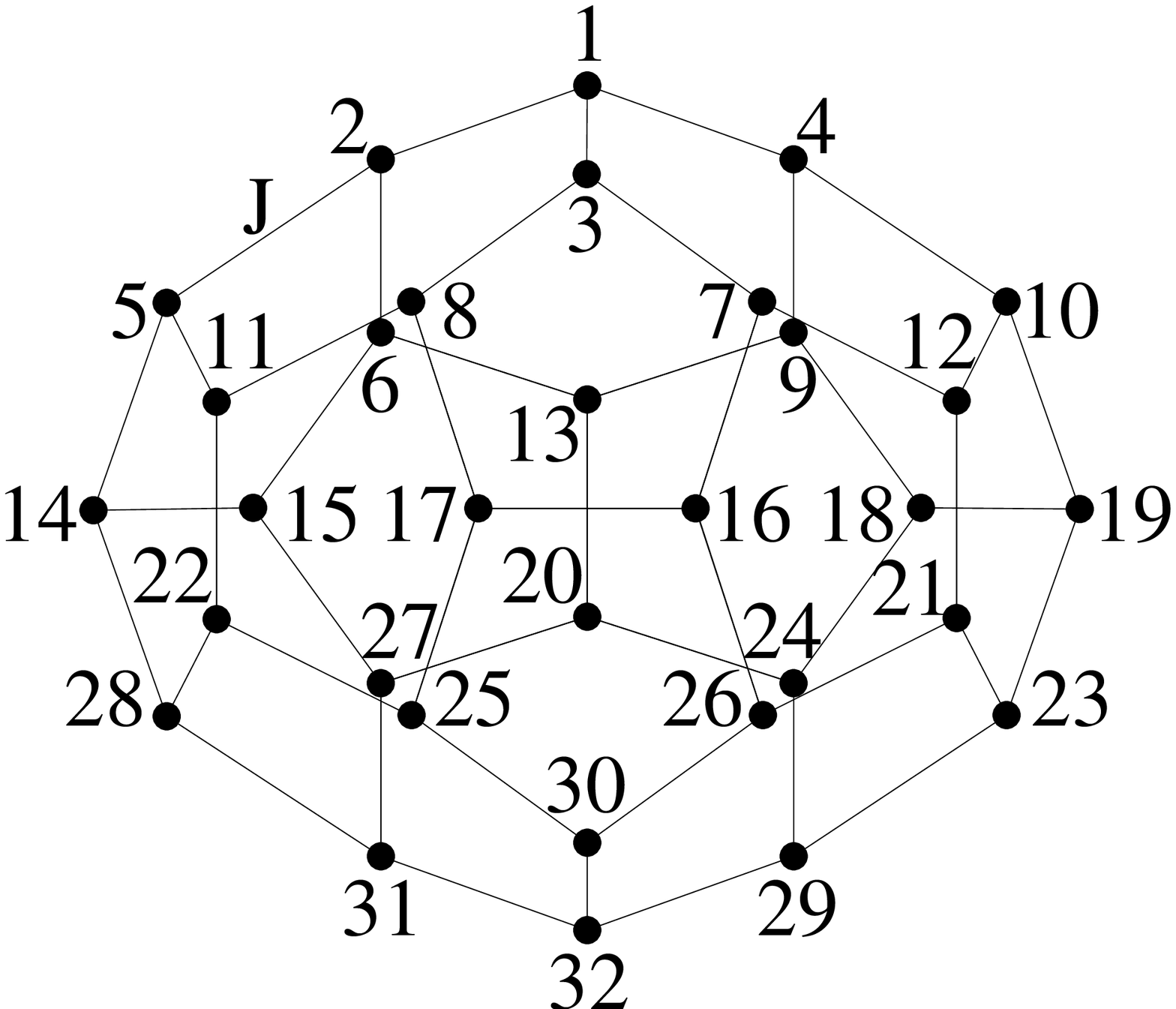}
\text{(e) \hspace{103pt} (f)}

\caption{Projection of the clusters on a plane: (a) $C_{24}$, (b) $C_{26}$, (c)
$C_{28}$, (d) $C_{30}$ (e) $C_{32,I}$ and (f) $C_{32,II}$. For $C_{30}$ there is
also a top (bottom with the dashes) view. The black circles are spins $s_{i}$.
The solid lines are antiferromagnetic bonds $J$.}
\label{fig:all}
\end{figure}

Motivated by the non-trivial spectral and magnetic properties of the AHM on the
$I_{h}$ clusters, the $s_{i}=\frac{1}{2}$ AHM on fullerene clusters with
symmetry other than $I_h$ is investigated.
Modine and Kaxiras calculated ground state energies and correlation functions
of clusters with $n$ up to 28, and by truncating the Hilbert space of
a cluster with $n=32$ \cite{Modine96}. They used no symmetries to reduce the
computational requirements and found that same hexagon correlations
are the most important for the minimization of the energy. For the $n=28$
cluster more specifically the ground state was found to be
doubly-degenerate. Their results are extended here
with the calculation of the low-energy excitations and the response in a
magnetic field for the clusters of Ref. \cite{Modine96} with $n=24$ and
28, the most symmetric for these $n$ values. The magnetic behavior of different
clusters with $n=26$ and 32 is investigated, as well as for a
cluster with $n=30$. All of them are shown in Fig. \ref{fig:all}
\cite{Fowler95,Yoshida97}. They are the most symmetric isomers for each $n$
(for $n=32$ the two clusters belong to different point groups who have
the same number of symmetry operations). The spatial symmetry group along with
the number of symmetry operations is listed in Table \ref{table:1} for each
$n$. The low energy
spectra are calculated with Lanczos diagonalization, and point-group and
spin-inversion symmetries are used to reduce the computational requirements and
to classify the states according to the total symmetry group's irreducible
representations \cite{NPK05-1}. Similarly to the $I_h$-symmetry case it is of
interest to look for non-magnetic states inside the
singlet-triplet gap, and search for
unconventional behavior of the magnetization and the possible presence of
discontinuities in a field. Comparing with the case of the $I_h$ clusters we
gauge the effect of symmetry on the behavior of the AHM on the fullerenes. It
is noted that fullerene clusters are edge- and not corner-sharing, and it is
not obvious if it is possible to re-write the Hamiltonian as a sum of total
spins on individual units or how to perform any other mathematical operations
in order to derive analytic results, even in the classical limit.

\begin{figure}
\includegraphics[width=6.5in,height=3.5in]{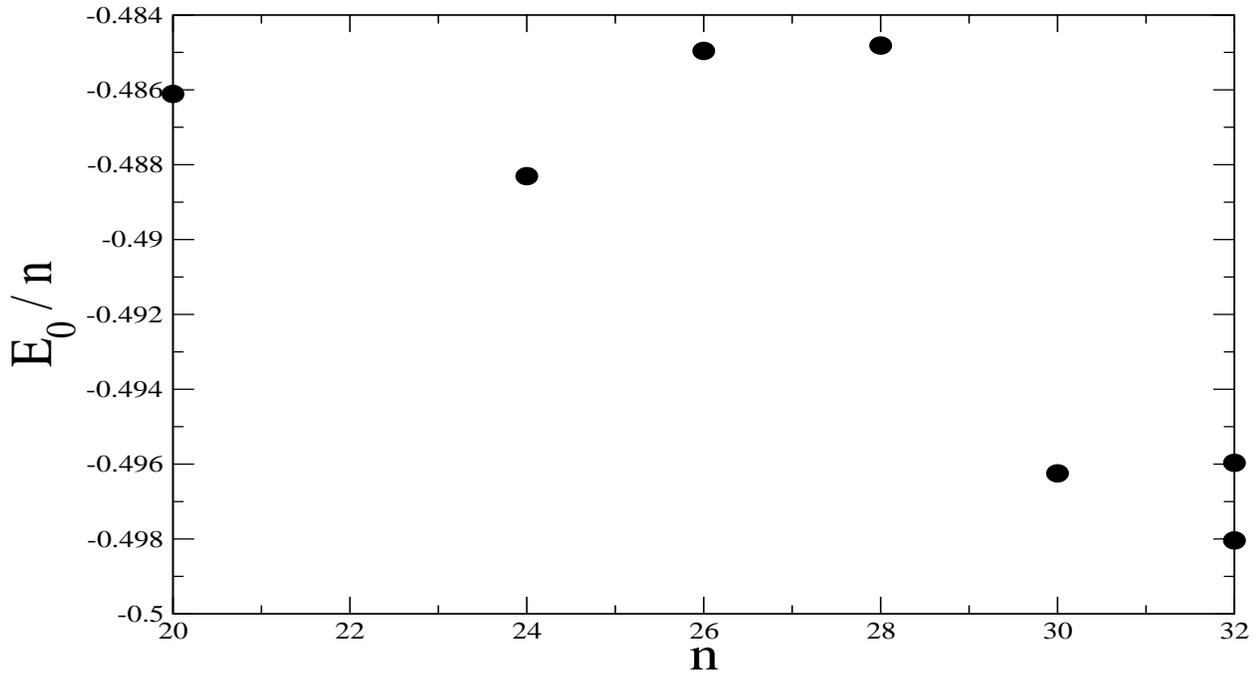}
\caption{Ground state energy per spin $\frac{E_0}{n}$ as a function of the
number of spins $n$. The value for $n=20$ is taken from \cite{NPK05-1}. The
lowest energy for $n=32$ is for cluster $C_{32,II}$, and the highest for
$C_{32,I}$.}
\label{fig:61}
\end{figure}

From now on sites which belong only to pentagons will be called pentagon sites,
while the rest will be called hexagon sites. The clusters will be abbreviated
as $C_n$, with the $D_{3d}$ $n=32$ cluster $C_{32,I}$, and the $D_{3h}$ $n=32$
cluster $C_{32,II}$. Looking at Fig. \ref{fig:all}, the pentagons form a band in
the middle of $C_{24}$, while the hexagons a band in the middle of $C_{30}$ and
$C_{32,I}$. For $C_{26}$ and $C_{28}$ no hexagons are sharing edges, while for
$C_{32,II}$ the hexagons are adjacent to each other in two groups of three.
Close proximity of polygons of the same kind can in principle minimize
frustration, which also decreases on the average with $n$. As seen in Table
\ref{table:1} and Fig. \ref{fig:61} the ground state energy per spin closely
follows the aforementioned arguments. The clusters with $n=30$ and 32 have the
lowest energy per spin, as the hexagons approach each other more and more
with increasing $n$. $C_{24}$ has the next lowest energy with the band of
pentagons separating the two hexagons.
For $C_{26}$ and $C_{28}$ with the disconnected hexagons the energy is maximum,
even higher than the dodecahedron ($C_{20}$) energy per spin, which is also
plotted for comparison \cite{NPK05-1}.

Frustration also results in non-magnetic excitations inside the singlet-triplet
gap. The first excited state in all cases except $C_{28}$ is a singly or doubly
degenerate singlet (Tables \ref{table:3}-\ref{table:8}, Fig. \ref{fig:60}). For
$C_{30}$ and $C_{32,I}$ the second excited state is also
a singlet, non-degenerate. In contrast, $C_{28}$'s first excited state is a
triply degenerate triplet. For $C_{24}$, $C_{30}$ and $C_{32,I}$ the low-lying
singlets are well separated from the magnetic excitations. For all the clusters
the dominant nearest-neighbor correlations in the ground states are of the
intra-hexagon type, with intra-pentagon following closely (Table
\ref{table:9}). The only exception is $C_{32,II}$ where the intra-pentagon
correlation is very strong and equal to -0.60548. This is appointed to the
geometry of the cluster, with the ``strong'' bonds little
frustrated in the middle of the structure and between the two groups of three
hexagons each.

The magnetization in an external magnetic field does not present a
discontinuity, as is the case for the dodecahedron \cite{NPK05-1}. There are
only magnetization plateaux, where the ground state belongs to specific total
spin $S$ sectors for a wider range of fields than their neighboring sectors
(Fig. \ref{fig:7}). The most pronounced case is for $n=28$ and $S=2$. It
appears that reducing the symmetry from $I_h$ works against magnetization
discontinuities.

The clusters considered in this paper have relatively small $n$ and mostly
belong to different spatial symmetry groups. Only $C_{26}$ and $C_{32,II}$
have the same spatial symmetry, neverthless they have no common pattern of low
energy and magnetic behavior as was the case for the $I_h$ symmetry clusters
\cite{NPK05-1,NPK07}. Similarities are only found in the low energy behavior
of $C_{30}$ and $C_{32,I}$ even though they belong to different symmetry
groups, and they only have in common all the hexagons forming a band in the
middle. It is also pointed out that there are some minor differences between
the type I and type II molecules as they were called in Ref. \cite{NPK07}, even
though they both share the icosahedral $I_h$ symmetry. Larger clusters can shed
light on the magnetic properties as a function of the distribution of the
pentagons, however this is not possible with present day computational means,
at least for the low energy properties. Neverthless, connectivity appears to be
as important as spatial symmetry for the magnetic properties.

The plan of this paper is as follows: in section \ref{sec:2} the model and the
method are introduced, and in section \ref{sec:3} the low energy spectra
and nearest-neighbor correlation functions are presented.
Section \ref{sec:4} presents the results on the ground state magnetization, and
section \ref{sec:5} the conclusions.

\section{Model and Method}
\label{sec:2}
The antiferromagnetic Heisenberg Hamiltonian with spins $\vec{s}_{i}$ located
on cluster vertices $i$ is
\begin{equation}
H = J \sum_{<i,j>} \textrm{} \vec{s}_{i} \cdot \vec{s}_{j} - h S^{z}
\label{eqn:1}
\end{equation}
where $<>$ denotes nearest neighbors, and $J$ is positive and will be set
equal to $1$ from now on, defining the unit of energy. $h$ is the strength of
an external magnetic field and $S^{z}$ the projection of the total spin along
the field direction $z$. For an explanation of the use of symmetry operations
to block-diagonalize the Hamiltonian, see Ref. \cite{NPK05-1} and references
therein. The data for the symmetry groups was taken from Ref. \cite{Altmann94}.
Degeneracies are reported with respect to states with specific $S$, each of
which corresponds to $2S+1$ states with different value of $S^z$. Lanczos
diagonalization was performed in double precision, but results are shown with a
smaller number of significant digits to facilitate the presentation.

\section{Low Energy Spectra and Correlation Functions}
\label{sec:3}

The ground state energies per spin along with their degeneracies are listed in
Table \ref{table:1}. They are plotted as a function of $n$ in Fig.
\ref{fig:61}. The bigger clusters achieve the lowest energy, while $C_{26}$ and
$C_{28}$ the highest. Proximity of polygons of the same kind is crucial for
energy minimization \cite{Modine96}. $C_{26}$ and $C_{28}$ do not have hexagons
adjacent to each other (Fig. \ref{fig:all}), and their energy per spin is even
higher than the corresponding energy for the dodecahedron, which has no
hexagons \cite{NPK05-1}. They show that increase of the number of hexagons
does not necessarily lead to lower energy per spin. For the bigger clusters the
hexagons approach each other more and more and lower the energy, with
$C_{32,II}$ having the lowest. Its structure is such that there are two groups
of three adjacent hexagons each (Fig. \ref{fig:all}(f)), and all
hexagon-hexagon bond correlations are strong
(Table \ref{table:9}). In addition, there is a set of three
pentagon-pentagon bonds in the middle of the cluster
that have a very strong singlet character with a correlation value equal to
-0.60548. Such a strong correlation is not achieved even within hexagons in
any of the clusters, and points to the importance of the specific structure
geometry rather than the symmetry for the minimization of the energy, at least
for the fullerene clusters with small $n$ considered in this paper.
The two hexagon groups and the strong singlet-like pentagon bonds
are correlated weakly with the rest of the spins.

\begin{figure}
\includegraphics[width=6.5in,height=3.5in]{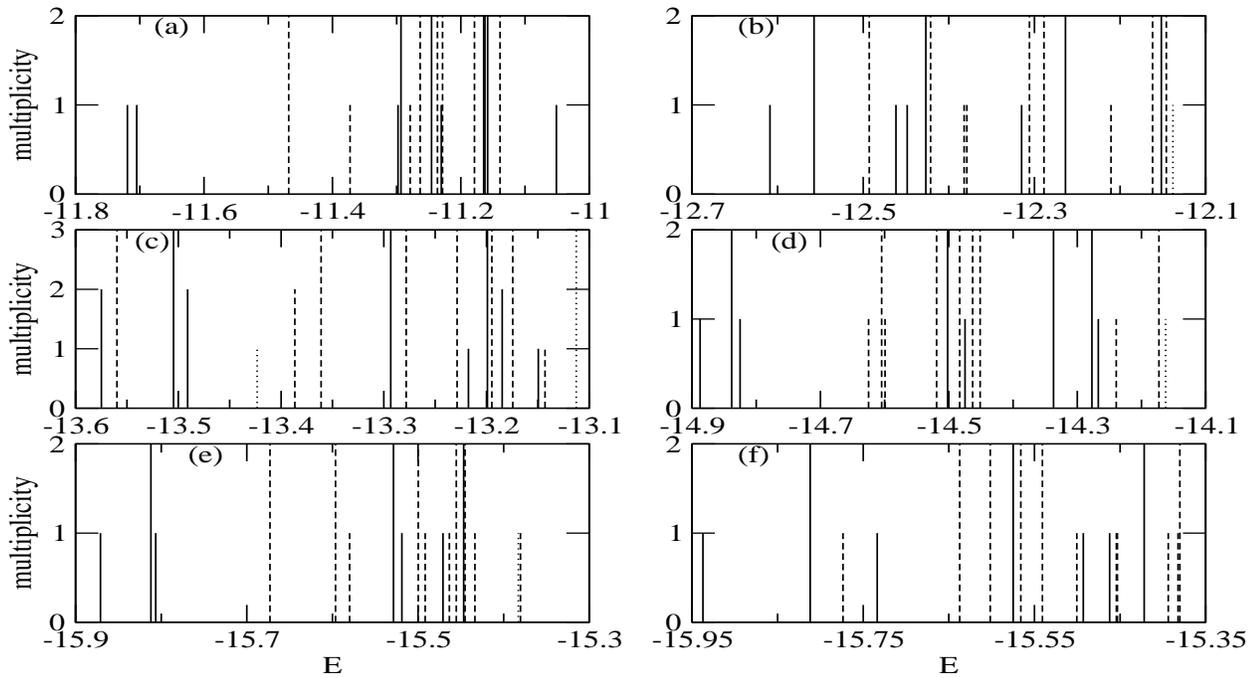}
\caption{Low energy $E$ spectrum of total spin $S$ states and its multiplicity:
(a) $C_{24}$, (b) $C_{26}$, (c) $C_{28}$, (d) $C_{30}$ (e) $C_{32,I}$ and (f)
$C_{32,II}$. Solid lines: $S=0$, dashed lines: $S=1$, dotted lines: $S=2$.}
\label{fig:60}
\end{figure}

The low energy spectra are shown in Tables \ref{table:3}-\ref{table:8} and
Fig. \ref{fig:60}. Except from $C_{28}$, the ground and first excited states
are singlets, with the ground states non-degenerate and the first excited
states doubly degenerate except from $C_{24}$ where it is non-degenerate. For
$C_{30}$ and $C_{32,I}$ the second excited state is also a singly degenerate
singlet. As seen in Figs. \ref{fig:60}(d) and \ref{fig:60}(e) the low energy
spectra of $C_{30}$ and $C_{32,I}$ are similar. There is a non-degenerate
singlet followed by two closely spaced singlets with the same degeneracy. The
two clusters belong to different symmetry groups, but they both have their
hexagons forming a band in the middle. Spin inversion symmetry is however
opposite for the six lowest states in the spectra (Tables \ref{table:6} and
\ref{table:7}). $C_{26}$ and $C_{32,II}$ on the other hand have the same spatial
symmetry, $D_{3h}$. In their low energy spectra
(Tables \ref{table:4} and \ref{table:8}) the ground states belong to
different irreducible representations, and their spin inversion symmetry
is also different. The first excited states belong to the same irreducible
representation, however the spin inversion symmetry is still different. This
is in contrast to the smallest dodecahedral $I_h$ symmetry clusters, the
icosahedron (not of the fullerene type but made only of triangles) and the
dodecahedron, which have similar
structure and relative spacing of the levels in their
low energy spectra even though they comprise of different polygons,
therefore symmetry is a strong determining factor for their properties. This
result points to the conclusion that spatial symmetry is not the only factor
determining the low energy properties in general for the fullerenes.

$C_{28}$ differs from the rest of the clusters in that its ground state is a
doubly degenerate singlet \cite{Modine96}, while the first excited state is a
closely spaced triply degenerate triplet (Table \ref{table:5}). Then triply and
doubly degenerate singlets follow, and the first non-degenerate state which has
$S=2$. In no other cluster an $S=2$ state lies so low in the excitation
spectrum. The ground state doublet belongs to the $E_{s}$ representation, which
transforms as the pair $(x^{2}-y^{2},2z^{2}-x^{2}-y^{2})$ of Cartesian tensors
\cite{Altmann94}. In contrast to \cite{Modine96}, here we find the correlations
to be the same for both degenerate ground states. The ground state nearest
neighbor correlation functions are shown in Table \ref{table:9} for all the
clusters. Except from the intra-pentagon correlation of $C_{32,II}$ same hexagon
correlations are the strongest.

The nearest-neighbor correlation functions for the lowest singlet excitations
are listed in table \ref{table:10}. The two smallest clusters and $C_{32,II}$
lower the energy of the hexagon-pentagon bonds while increasing the energy of
the rest (only one of the same-hexagon bonds of $C_{26}$ lowers very weakly).
On the contrary, $C_{30}$ behaves the other way, except from the intra-hexagon
correlation in the middle of the cluster that refers to a common side of two
hexagons, which weakens. Its two singlet excited states are very close in
energy and in the behavior of the correlation functions, even though they are
of different multiplicity, which is also true for $C_{32,I}$. The latter alters
its same-hexagon bonds to generate the two excited singlets, even though the
strongest one that refers to a common side of two hexagons does not change
significantly.

The nearest-neighbor correlation functions for the first triplet excitation are
shown in Table \ref{table:11}. The values for $C_{32,II}$ change relatively
little compared to the ground state, and the pentagon-pentagon correlation is
getting even stronger. Similarly, $C_{30}$ shows little change except from the
hexagon-pentagon correlation that gets weaker. For $C_{24}$ the triplet is
mostly due to the decrease of the intra-hexagon correlation function, while in
$C_{26}$ the inter-hexagon correlation is getting stronger. In $C_{28}$ only
the inter-hexagon correlation $<\vec{S}_1 \cdot \vec{S}_5>$ changes relatively
weakly compared to the other correlations, and correlation
$<\vec{S}_{10} \cdot \vec{S}_{11}>$ is particularly weak. Finally, for
$C_{32,I}$ there are strong changes for all correlations except from
$<\vec{S}_5 \cdot \vec{S}_{16}>$.

\section{Ground State Magnetization}
\label{sec:4}

The ground state magnetization as a function of an external field has typically
a step-like structure, with a $\Delta S=1$ discontinuity at fields
where the ground state switches between adjacent $S$ sectors.
Frustration though can lead to magnetization discontinuities with $\Delta
S>1$, where a particular $S$ sector never necomes the ground state in a
field. Such is the case for the icosahedral symmetry $I_{h}$ clusters, where
the sector $S=\frac{n}{2}-5$ with five flipped spins from saturation never
includes the ground state \cite{NPK07}. The number of
discontinuities is more than one at the classical level $S \to \infty$, and
also for the dodecahedron ($n=20$) and $s_{i}=1$ where the calculation of the
lowest energy state is computationally feasible for all $S$ sectors (for $I_h$
clusters with $n > 20$ the lowest energy state calculation is only possible for
very high $S$ even for $s_{i}=\frac{1}{2}$).

\begin{figure}
\includegraphics[width=6.5in,height=3.5in]{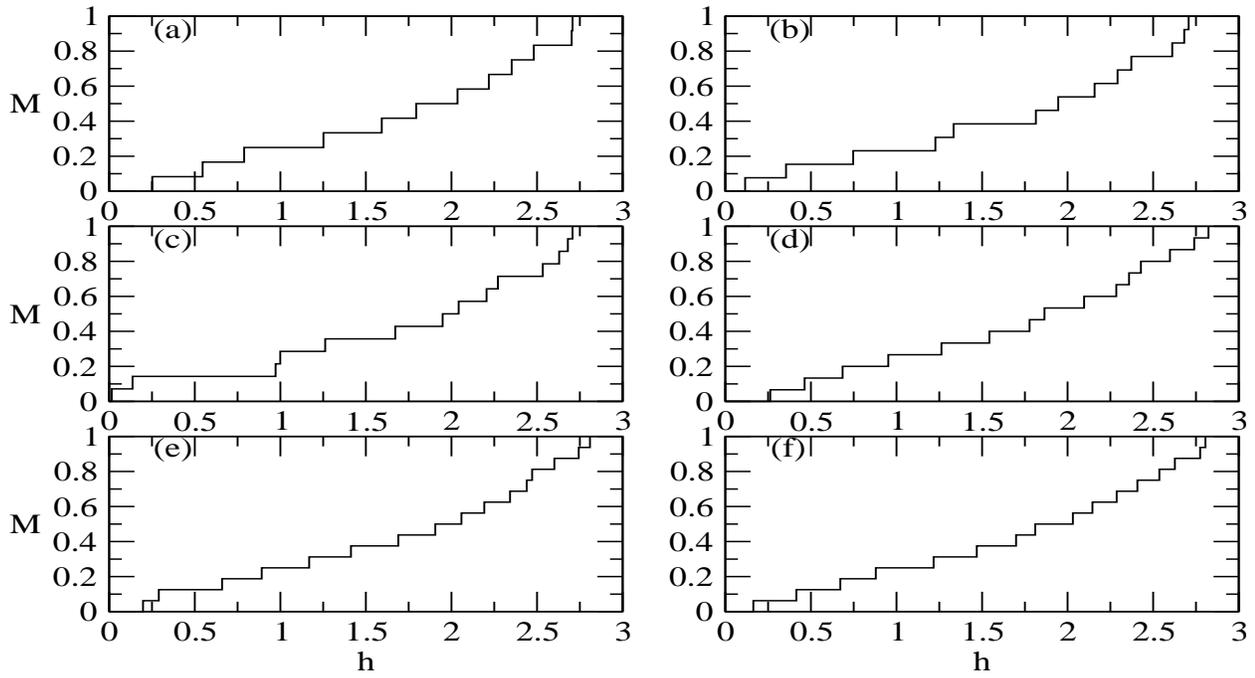}
\caption{Reduced ground state magnetization $M=\frac{S}{n s_{i}}$ as a function
of magnetic field $h$: (a) $C_{24}$, (b) $C_{26}$, (c) $C_{28}$, (d) $C_{30}$, (e)
$C_{32,I}$ and (f) $C_{32,II}$. $M$ is the total spin $S$ normalized to the
number of sites $n$ and the magnitude of spin $s_i$. $M$ has no units and $h$
is in units of energy.}
\label{fig:7}
\end{figure}

The lowest states for all the $S$ sectors along with their degeneracies and the
irreducible representation to which they belong are listed in Table
\ref{table:12} (the saturation fields are listed in Table \ref{table:1}).
The corresponding reduced magnetization $M=\frac{S}{n s_i}$
curves are shown in Fig. \ref{fig:7}. Unlike the icosahedral symmetry case, no
discontinuities are found.

For some values of $M$ there are plateaux, where a particular $S$ sector
contains the ground state for a wider range of fields than the neighboring
sectors. The most pronounced appears for $C_{28}$ and $S=2$, where $M=0.14286$
(Fig. \ref{fig:7}(c))).

\begin{figure}
\includegraphics[width=6.5in,height=3.5in]{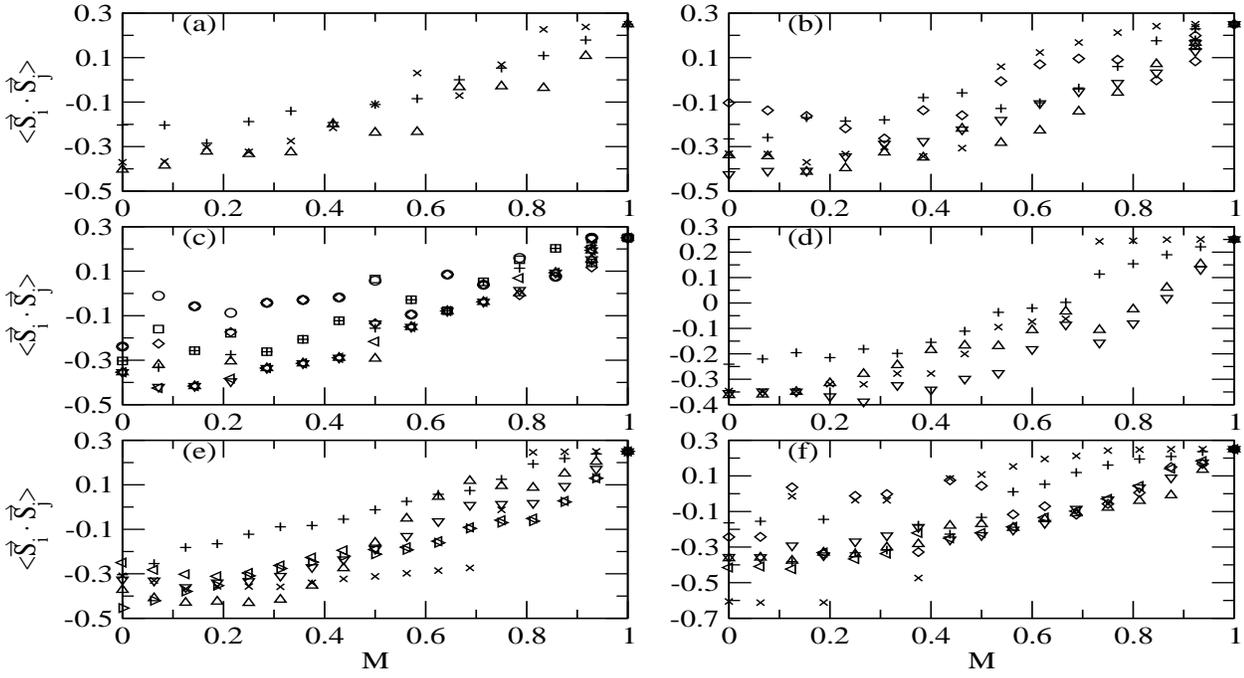}
\caption{Distinct correlation functions for the lowest energy state in each
total spin $S$ sector: (a) $C_{24}$, (b) $C_{26}$, (c) $C_{28}$, (d) $C_{30}$, (e)
$C_{32,I}$ and (f) $C_{32,II}$. The reduced magnetization $M=\frac{S}{n s_{i}}$
is the total spin $S$ normalized to the number of sites $n$ and the magnitude
of spin $s_i$. $<\vec{S}_{i} \cdot \vec{S}_{j}>$ is in units of energy and $M$
has no units. $<\vec{S}_{i} \cdot \vec{S}_{j}>$: $\vartriangle$, $\lhd$, $\rhd$,
$\triangledown$: intra-hexagon, $\diamond$, $\circ$: inter-hexagon, +,
$\square$: hexagon-pentagon, $\times$: intra-pentagon.}
\label{fig:8}
\end{figure}

It is again hard to draw correlations between symmetry and the response in a
magnetic field. In Figs. \ref{fig:7}(a) and \ref{fig:8}(a) there is a
correlation of the plateau-like features of the magnetization curve of $C_{24}$
with stronger values of the intra-hexagon bonds. For $C_{28}$, where the
singlet-triplet gap is very small, there are
stronger intra-hexagon correlation functions for the few low-lying $S>0$
sectors compared to the singlet case (Fig. \ref{fig:8}(c)). There are $S$
sectors that contain the ground state for a very narrow range of the field
(Fig. \ref{fig:7}(c)), and looking at Table \ref{table:12} their lowest state
belongs to three-dimensional irreducible representations. For $C_{30}$, sectors
$S=3$ and $4$ have very strong same-hexagon correlations, stronger than the
ones in the ground state (Fig. \ref{fig:8}(d)). For $C_{32,I}$ there are strong
same-hexagon correlations for the low-$S$ sectors, but for higher $S$
intra-pentagon correlations are the strongest (Fig. \ref{fig:8}(e)). In both
cases, the strength of these correlations does not change significantly with
the $S$ value. $S=12$ is the first sector that restores the same-hexagon
correlations as the strongest, and it is the ground state for a narrow
field window (Fig. \ref{fig:7}(e)). In the case of $C_{32,II}$ there are low-$S$
sectors where the intra-pentagon correlation is very strong (Fig.
\ref{fig:8}(f)). For $S=1$ and $3$ it is even stronger than the $S=0$ value.
For $C_{26}$ the $S=4$ and $6$ sectors are ground states for a narrow range of
the field (Fig. \ref{fig:7}(b)) and same-hexagon correlations are weak (Fig.
\ref{fig:8}(b)). Finally, for $C_{24}$ the single spin-flip subspace has the
ground state for a very narrow window of the field (Fig. \ref{fig:7}(a)), with
the two spin-flip subspace having a plateau and the strongest intra-hexagon
correlations relative to its neighboring $S$ sectors (Fig. \ref{fig:8}(a)).

For $C_{26}$ and $C_{28}$ the sector with a single spin flip from saturation is
degenerate (Table \ref{table:12}). However, the spin flips are not confined on
the hexagons except from the singly degenerate state of $C_{26}$, therefore
there is no analog with the high magnetization localized magnon states
discussed in Ref. \cite{Schulenburg02}.

\section{Conclusions}
\label{sec:5}

The low energy spectrum and the magnetic response of the AHM have been
calculated on a series of clusters with the connectivity of the fullerenes and
a number of sites ranging from $24$ to $32$. Frustration and connectivity have
a signature on the low energy spectrum with singlet ground and low energy
excited states, the only exception being the $28$-site cluster where the
ground state is a doubly degenerate singlet and the first excited state a
triplet. Frustration is minimal when pentagons and hexagons minimize their
interference in the clusters by being placed adjacent to polygons of the same
kind. The magnetization as a function of an external field exhibits plateaux
features, the most pronounced for $n=28$ and $S=2$. Unlike the icosahedral
$I_h$ symmetry clusters \cite{NPK05-1,NPK07}, spatial symmetry is not the sole
determining factor of the magnetic properties of the clusters and their
connectivity appears to be important as well. It is desirable to investigate
clusters of higher $n$ to gain more insight on the correlation between
symmetry, connectivity and magnetic properties, however this is very
challenging with present day computational means, at least for the low energy
properties. For low $n$ the competition between unfrustrated hexagons and
frustrated pentagons is strong, but even for high $n$ the pentagon influence
can be important, as in the case of $I_h$ symmetry \cite{NPK07}.

The author thanks D. Coffey for discussions. Most of the calculations were
carried out at the Trinity Center for High Performance Computing at the
University of Dublin. The work was supported by a Marie Curie Fellowship of the
European Community program Development Host Fellowship under contract number
HPMD-CT-2000-00048.

$^{*}$ Present address: Institut f\"ur Theoretische Physik A, Physikzentrum,
RWTH Aachen, 52056 Aachen, Germany, Institut f\"ur
Festk\"orperforschung-Theorie III, Forschungszentrum J\"ulich,
Leo-Brandt-Strasse, 52425 J\"ulich, Germany and JARA-Fundamentals of Future
Information Technology.

\bibliography{paperfourpointfive}

%
%

\begin{table}[h]
\begin{center}
\caption{Symmetry and ground state properties for the five clusters.
$\frac{E_0}{n}$ is the ground state energy per spin and $mult.$ is the state's
multiplicity. $h_{sat}$ is the saturation field.}
\begin{tabular}{c|c|c|c|c|c}
sites & symmetry & number of & $\frac{E_0}{n}$ & $mult.$ &$h_{sat}$\\
$n$ & group & symmetry & & & \\
& & operations & & & \\
\hline
24      & $D_{6d}$ & 24 & -0.48831 & 1 & 4 + $\sqrt{2}$ \\
\hline
26      & $D_{3h}$ & 12 & -0.48496 & 1 & 4 + $\sqrt{2}$ \\
\hline
28      & $T_{d}$  & 24 & -0.48482 & 2 & 4 + $\sqrt{2}$ \\
\hline
30      & $D_{5h}$ & 20 & -0.49625 & 1 & 3 + $\sqrt{7}$ \\
\hline
32 (I)  & $D_{3d}$ & 12 & -0.49597 & 1 & $\frac{9+\sqrt{5}}{2}$ \\
\hline
32 (II) & $D_{3h}$ & 12 & -0.49804 & 1 & 5.61050 \\
\end{tabular}
\label{table:1}
\end{center}
\end{table}

\begin{table}[h]
\begin{center}
\caption{Low energy spectrum for $C_{24}$. $E$ is the energy, $mult.$ stands for
the multiplicity of the state and $irrep.$ for irreducible representation. $S$
is the total spin, with each $S$ state corresponding to $2S+1$ states with
different projection of the total spin along the $z$ axis $S^z$. The spatial
irreducible representation notation follows Ref. \cite{Altmann94}. Indices $s$
and $a$ indicate the behavior under spin inversion, where $s$ stands for
symmetric and $a$ for antisymmetric. A comma is introduced when necessary to
avoid confusion between the notation for the spatial irreducible
representation and the behavior under spin inversion.}
\begin{tabular}{c|c|c|c|c|c|c|c}
$E$ & mult. & irrep. & $S$ & $E$ & mult. & irrep. & $S$ \\
\hline
-11.71937 & 1 & $B_{1,s}$ & 0 & -11.23652 & 2 & $E_{2,a}$ & 1 \\
\hline
-11.70478 & 1 & $A_{1,s}$ & 0 & -11.23043 & 1 & $A_{1,s}$ & 0 \\
\hline
-11.46814 & 2 & $E_{3,a}$ & 1 & -11.22852 & 2 & $E_{1,a}$ & 1 \\
\hline
-11.37244 & 1 & $A_{2,a}$ & 1 & -11.17883 & 2 & $E_{3,a}$ & 1 \\
\hline
-11.29779 & 1 & $B_{1,s}$ & 0 & -11.16437 & 2 & $E_{4,s}$ & 0 \\
\hline
-11.29326 & 2 & $E_{2,s}$ & 0 & -11.16257 & 2 & $E_{1,s}$ & 0 \\
\hline
-11.27888 & 1 & $B_{2,a}$ & 1 & -11.15826 & 2 & $E_{3,s}$ & 0 \\
\hline
-11.26341 & 2 & $E_{5,a}$ & 1 & -11.13896 & 2 & $E_{4,a}$ & 1 \\
\hline
-11.24565 & 2 & $E_{5,s}$ & 0 & -11.05126 & 1 &  $A_{1,s}$ & 0 \\
\end{tabular}
\label{table:3}
\end{center}
\end{table}

\begin{table}
\begin{center}
\caption{Low energy spectrum for $C_{26}$. Notation as in table \ref{table:3}.}
\begin{tabular}{c|c|c|c|c|c|c|c}
$E$ & mult. & irrep. & $S$ & $E$ & mult. & irrep. & $S$ \\
\hline
-12.60898 & 1 & $A_{2,a}^{'}$  & 0 & -12.31502 & 1 & $A_{2,a}^{'}$  & 0 \\
\hline
-12.55739 & 2 & $E_{a}^{'}$    & 0 & -12.30577 & 2 & $E_{s}^{''}$  & 1 \\
\hline
-12.49297 & 2 & $E_{s}^{'}$    & 1 & -12.28884 & 2 & $E_{s}^{'}$   & 1 \\
\hline
-12.46174 & 1 & $A_{2,a}^{''}$ & 0 & -12.26375 & 2 & $E_{a}^{''}$  & 0 \\
\hline
-12.44862 & 1 & $A_{2,a}^{''}$ & 0 & -12.21057 & 1 & $A_{1,s}^{'}$  & 1 \\
\hline
-12.42682 & 2 & $E_{a}^{'}$    & 0 & -12.16192 & 2 & $E_{s}^{''}$ & 1 \\
\hline
-12.42129 & 2 & $E_{s}^{'}$    & 1 & -12.15183 & 2 & $A_{1,a}^{''}$ & 0 \\
\hline
 -12.38217 & 1 & $A_{1,s}^{'}$  & 1 & -12.14598 & 2 & $E_{s}^{'}$ & 1 \\
\hline
 -12.37889 & 1 & $A_{1,s}^{''}$ & 1 & -12.13819 & 1 & $A_{2,a}^{'}$ & 2 \\
\end{tabular}
\label{table:4}
\end{center}
\end{table}

\begin{table}
\begin{center}
\caption{Low energy spectrum for $C_{28}$. Notation as in table \ref{table:3}.}
\begin{tabular}{c|c|c|c|c|c|c|c}
$E$ & mult. & irrep. & $S$ & $E$ & mult. & irrep. & $S$ \\
\hline
-13.57486 & 2 & $E_{s}$  & 0 & -13.22866 & 3 & $T_{2,a}$ & 1 \\
\hline
-13.55978 & 3 & $T_{2,a}$ & 1 & -13.21762 & 1 & $A_{1,s}$ & 0 \\
\hline
-13.50468 & 3 & $T_{2,s}$ & 0 & -13.19914 & 3 & $T_{2,s}$ & 0 \\
\hline
-13.49099 & 2 & $E_{s}$  & 0 & -13.19486 & 3 & $T_{1,a}$ & 1 \\
\hline
-13.42327 & 1 & $A_{1,s}$ & 2 & -13.18471 & 2 & $E_{s}$  & 0 \\
\hline
-13.38652 & 2 & $E_{a}$  & 1 & -13.17446 & 3 & $T_{2,a}$ & 1 \\
\hline
-13.36105 & 3 & $T_{2,a}$ & 1 & -13.14963 & 1 & $A_{2,s}$ & 0 \\
\hline
-13.29319 & 3 & $T_{1,s}$ & 0 & -13.14309 & 1 & $A_{1,a}$ & 1 \\
\hline
-13.27818 & 3 & $T_{1,a}$ & 1 & -13.11265 & 3 &  $T_{2,s}$ & 2 \\
\end{tabular}
\label{table:5}
\end{center}
\end{table}

\begin{table}
\begin{center}
\caption{Low energy spectrum for $C_{30}$. Notation as in table \ref{table:3}.}
\begin{tabular}{c|c|c|c|c|c|c|c}
$E$ & mult. & irrep. & $S$ & $E$ & mult. & irrep. & $S$ \\
\hline
-14.88742 & 1 & $A_{2,a}^{''}$ & 0 & -14.47476 & 1 & $A_{2,a}^{''}$ & 0 \\
\hline
-14.83815 & 2 & $E_{2,a}^{''}$ & 0 & -14.46287 & 2 & $E_{2,s}^{''}$ & 1 \\
\hline
-14.82517 & 1 & $A_{2,a}^{'}$  & 0 & -14.45130 & 2 & $E_{1,s}^{''}$ & 1 \\
\hline
-14.62495 & 1 & $A_{1,s}^{'}$  & 1 & -14.33694 & 2 & $E_{1,a}^{'}$  & 0 \\
\hline
-14.60458 & 2 & $E_{2,s}^{'}$  & 1 & -14.27711 & 2 & $E_{1,a}^{''}$ & 0 \\
\hline
-14.59928 & 1 & $A_{1,s}^{''}$ & 1 & -14.26727 & 1 & $A_{2,a}^{'}$  & 0 \\
\hline
-14.51907 & 2 & $E_{1,s}^{'}$  & 1 & -14.23932 & 1 & $A_{1,s}^{'}$  & 1 \\
\hline
-14.50190 & 2 & $E_{1,a}^{''}$ & 0 & -14.17268 & 2 & $E_{2,s}^{''}$ & 1 \\
\hline
-14.48316 & 2 & $E_{2,s}^{'}$  & 1 & -14.16221 & 1 & $A_{2,a}^{'}$  & 2 \\
\end{tabular}
\label{table:6}
\end{center}
\end{table}

\begin{table}
\begin{center}
\caption{Low energy spectrum for $C_{32,I}$. Notation as in table
\ref{table:3}.}
\begin{tabular}{c|c|c|c|c|c|c|c}
$E$ & mult. & irrep. & $S$ & $E$ & mult. & irrep. & $S$ \\
\hline
-15.87092 & 1 & $A_{1u,s}$ & 0 & -15.49168 & 1 & $A_{2g,a}$ & 1 \\
\hline
-15.81199 & 2 & $E_{g,s}$  & 0 & -15.47080 & 1 & $A_{1u,s}$ & 0 \\
\hline
-15.80648 & 1 & $A_{1g,s}$ & 0 & -15.46339 & 1 & $A_{1u,a}$ & 1 \\
\hline
-15.67299 & 2 & $E_{u,a}$  & 1 & -15.45531 & 2 & $E_{g,a}$  & 1 \\
\hline
-15.59634 & 2 & $E_{g,a}$  & 1 & -15.44665 & 2 & $E_{u,s}$  & 0 \\
\hline
-15.57987 & 1 & $A_{2u,a}$ & 1 & -15.44513 & 2 & $E_{g,a}$  & 1 \\
\hline
-15.52875 & 2 & $E_{g,s}$  & 0 & -15.43356 & 1 & $A_{2g,a}$ & 1 \\
\hline
-15.51890 & 1 & $A_{1g,s}$ & 0 & -15.38287 & 1 & $A_{1g,s}$ & 2 \\
\hline
-15.49978 & 2 & $E_{u,a}$  & 1 & -15.38034 & 1 & $A_{2u,a}$ & 1 \\
\end{tabular}
\label{table:7}
\end{center}
\end{table}

\begin{table}
\begin{center}
\caption{Low energy spectrum for $C_{32,II}$. Notation as in table
\ref{table:3}.}
\begin{tabular}{c|c|c|c|c|c|c|c}
$E$ & mult. & irrep. & $S$ & $E$ & mult. & irrep. & $S$ \\
\hline
-15.93723 & 1 & $A_{1,s}^{''}$ & 0 & -15.50045 & 1 & $A_{2,a}^{''}$ & 1 \\
\hline
-15.81192 & 2 & $E_{s}^{'}$    & 0 & -15.49288 & 1 & $A_{1,s}^{''}$ & 0 \\
\hline
-15.77366 & 1 & $A_{2,a}^{'}$  & 1 & -15.46219 & 1 & $A_{1,s}^{''}$ & 0 \\
\hline
-15.73368 & 1 & $A_{1,s}^{'}$  & 0 & -15.45437 & 1 & $A_{2,a}^{'}$ & 1 \\
\hline
-15.63730 & 2 & $E_{a}^{'}$    & 1 & -15.45317 & 1 & $A_{1,a}^{'}$ & 1 \\
\hline
-15.60167 & 2 & $E_{a}^{'}$    & 1 & -15.42185 & 2 & $E_{s}^{''}$ & 0 \\
\hline
-15.57485 & 2 & $E_{s}^{''}$   & 0 & -15.39363 & 1 & $A_{2,a}^{''}$ & 1 \\
\hline
-15.56589 & 2 & $E_{a}^{''}$   & 1 & -15.38231 & 1 & $A_{2,a}^{''}$ & 1 \\
\hline
-15.54070 & 2 & $E_{a}^{''}$   & 1 & -15.38020 & 2 & $E_{a}^{''}$ & 1 \\
\end{tabular}
\label{table:8}
\end{center}
\end{table}

\begin{table}
\begin{center}
\caption{Distinct nearest-neighbor correlation functions for the ground
states.}
\begin{tabular}{c|c|c|c|c}
sites $n$ & intra-hexagon & inter-hexagon & hexagon-pentagon & intra-pentagon \\
\hline
 24 & $<\vec{S}_1 \cdot \vec{S}_2> = -0.40325$ & & $<\vec{S}_1 \cdot \vec{S}_7> = -0.20285$ & $<\vec{S}_7 \cdot \vec{S}_8> = -0.37051$ \\
\hline
 26 & $<\vec{S}_5 \cdot \vec{S}_7> = -0.33858$ & $<\vec{S}_{11} \cdot \vec{S}_{12}> = -0.10317$ & $<\vec{S}_2 \cdot \vec{S}_5> = -0.26523$ & $<\vec{S}_1 \cdot \vec{S}_2> = -0.33215$ \\
    & $<\vec{S}_5 \cdot \vec{S}_{11}> = -0.42436$ & & & \\
\hline
 28 & $<\vec{S}_1 \cdot \vec{S}_2> = -0.35418$ & $<\vec{S}_1 \cdot \vec{S}_5>
 = -0.23883$ & $<\vec{S}_2 \cdot \vec{S}_3> = -0.30347$ & \\
\hline
 30 & $<\vec{S}_6 \cdot \vec{S}_7> = -0.36221$ &  & $<\vec{S}_1 \cdot \vec{S}_6> = -0.24081$ &  $<\vec{S}_1 \cdot \vec{S}_2> = -0.34618$ \\
 & $<\vec{S}_7 \cdot \vec{S}_{18}> = -0.35466$ & & & \\
\hline
 32 ($D_{3d}$) & $<\vec{S}_5 \cdot \vec{S}_6> = -0.37205$ &  & $<\vec{S}_2 \cdot \vec{S}_6> = -0.30381$ & $<\vec{S}_1 \cdot \vec{S}_2> = -0.30903$ \\
 & $<\vec{S}_5 \cdot \vec{S}_{16}> = -0.32629$ & & & \\
 & $<\vec{S}_{11} \cdot \vec{S}_{12}> = -0.24979$ & & & \\
 & $<\vec{S}_{11} \cdot \vec{S}_{19}> = -0.45409$ & & & \\
\hline
 32 ($D_{3h}$) & $<\vec{S}_1 \cdot \vec{S}_2> = -0.36273$ & $<\vec{S}_{11} \cdot \vec{S}_{22}> = -0.24352$ & $<\vec{S}_5 \cdot \vec{S}_{14}> = -0.16367$ & $<\vec{S}_{14} \cdot \vec{S}_{15}> = -0.60548$ \\
 & $<\vec{S}_2 \cdot \vec{S}_5> = -0.35602$ &  &  & \\
 & $<\vec{S}_5 \cdot \vec{S}_{11}> = -0.41480$ &  &  & \\
\end{tabular}
\label{table:9}
\end{center}
\end{table}

\begin{table}
\begin{center}
\caption{Distinct nearest-neighbor correlation functions for the singlet
excited states within the singlet-triplet gap.}
\begin{tabular}{c|c|c|c|c}
sites $n$ & intra-hexagon & inter-hexagon & hexagon-pentagon & intra-pentagon \\
\hline
 24 & $<\vec{S}_1 \cdot \vec{S}_2> = -0.37785$ & & $<\vec{S}_1 \cdot \vec{S}_7> = -0.25586$ & $<\vec{S}_7 \cdot \vec{S}_8> = -0.34169$ \\
\hline
 26 & $<\vec{S}_5 \cdot \vec{S}_7> = -0.28615$ & $<\vec{S}_{11} \cdot \vec{S}_{12}> = -0.084111$ & $<\vec{S}_2 \cdot \vec{S}_5> = -0.30251$ & $<\vec{S}_1 \cdot \vec{S}_2> = -0.30687$ \\
    & $<\vec{S}_5 \cdot \vec{S}_{11}> = -0.42640$ & & & \\
\hline
 30 & $<\vec{S}_6 \cdot \vec{S}_7> = -0.37609$ &  & $<\vec{S}_1 \cdot \vec{S}_6> = -0.21102$ &  $<\vec{S}_1 \cdot \vec{S}_2> = -0.35548$ \\
 & $<\vec{S}_7 \cdot \vec{S}_{18}> = -0.33026$ & & & \\
\hline
 30 & $<\vec{S}_6 \cdot \vec{S}_7> = -0.37784$ &  & $<\vec{S}_1 \cdot \vec{S}_6> = -0.20832$ &  $<\vec{S}_1 \cdot \vec{S}_2> = -0.35598$ \\
 & $<\vec{S}_7 \cdot \vec{S}_{18}> = -0.32507$ & & & \\
\hline
 32 ($D_{3d}$) & $<\vec{S}_5 \cdot \vec{S}_6> = -0.39744$ &  & $<\vec{S}_2 \cdot \vec{S}_6> = -0.30566$ & $<\vec{S}_1 \cdot \vec{S}_2> = -0.29805$ \\
 & $<\vec{S}_5 \cdot \vec{S}_{16}> = -0.29843$ & & & \\
 & $<\vec{S}_{11} \cdot \vec{S}_{12}> = -0.27220$ & & & \\
 & $<\vec{S}_{11} \cdot \vec{S}_{19}> = -0.45947$ & & & \\
\hline
 32 ($D_{3d}$) & $<\vec{S}_5 \cdot \vec{S}_6> = -0.40407$ &  & $<\vec{S}_2 \cdot \vec{S}_6> = -0.29314$ & $<\vec{S}_1 \cdot \vec{S}_2> = -0.30482$ \\
 & $<\vec{S}_5 \cdot \vec{S}_{16}> = -0.30669$ & & & \\
 & $<\vec{S}_{11} \cdot \vec{S}_{12}> = -0.27231$ & & & \\
 & $<\vec{S}_{11} \cdot \vec{S}_{19}> = -0.45357$ & & & \\
\hline
 32 ($D_{3h}$) & $<\vec{S}_1 \cdot \vec{S}_2> = -0.36028$ & $<\vec{S}_{11} \cdot \vec{S}_{22}> = -0.16865$ & $<\vec{S}_5 \cdot \vec{S}_{14}> = -0.24184$ & $<\vec{S}_{14} \cdot \vec{S}_{15}> = -0.40508$ \\
 & $<\vec{S}_2 \cdot \vec{S}_5> = -0.34068$ &  &  & \\
 & $<\vec{S}_5 \cdot \vec{S}_{11}> = -0.41156$ &  &  & \\
\end{tabular}
\label{table:10}
\end{center}
\end{table}

\begin{table}
\begin{center}
\caption{Distinct nearest-neighbor correlation functions for the first triplet
excited states.}
\begin{tabular}{c|c|c|c|c}
sites $n$ & intra-hexagon & inter-hexagon & hexagon-pentagon & intra-pentagon \\
\hline
 24 & $<\vec{S}_1 \cdot \vec{S}_2> = -0.38523$ & & $<\vec{S}_1 \cdot \vec{S}_7> = -0.20358$ & $<\vec{S}_7 \cdot \vec{S}_8> = -0.36686$ \\
\hline
 26 & $<\vec{S}_5 \cdot \vec{S}_7> = -0.34294$ & $<\vec{S}_{11} \cdot \vec{S}_{12}> = -0.13733$  & $<\vec{S}_2 \cdot \vec{S}_5> = -0.25900$ & $<\vec{S}_1 \cdot \vec{S}_2> = -0.33365$ \\
 & $<\vec{S}_5 \cdot \vec{S}_{11}> = -0.40945$ & & & \\
\hline
 28 & $<\vec{S}_1 \cdot \vec{S}_2> = -0.32073$ & $<\vec{S}_1 \cdot \vec{S}_5>
 = -0.22540$ & $<\vec{S}_2 \cdot \vec{S}_3> = -0.33309$ & \\
 & $<\vec{S}_4 \cdot \vec{S}_5> = -0.42032$ & $<\vec{S}_{10} \cdot \vec{S}_{11}> = -0.010665$ & $<\vec{S}_3 \cdot \vec{S}_4> = -0.16012$ & \\
 & $<\vec{S}_2 \cdot \vec{S}_{10}> = -0.42541$ & & & \\
\hline
 30 & $<\vec{S}_6 \cdot \vec{S}_7> = -0.35985$ & & $<\vec{S}_1 \cdot \vec{S}_6> = -0.22022$ & $<\vec{S}_1 \cdot \vec{S}_2> = -0.34857$ \\
 & $<\vec{S}_7 \cdot \vec{S}_{18}> = -0.34801$ & & & \\
\hline
 32 ($D_{3d}$) & $<\vec{S}_5 \cdot \vec{S}_6> = -0.40953$ &  & $<\vec{S}_2 \cdot \vec{S}_6> = -0.25415$ & $<\vec{S}_1 \cdot \vec{S}_2> = -0.32989$ \\
 & $<\vec{S}_5 \cdot \vec{S}_{16}> = -0.33089$ & & & \\
 & $<\vec{S}_{11} \cdot \vec{S}_{12}> = -0.28203$ & & & \\
 & $<\vec{S}_{11} \cdot \vec{S}_{19}> = -0.42064$ & & & \\
\hline
 32 ($D_{3h}$) & $<\vec{S}_1 \cdot \vec{S}_2> = -0.35863$ & $<\vec{S}_{11} \cdot \vec{S}_{22}> = -0.24266$ & $<\vec{S}_5 \cdot \vec{S}_{14}> = -0.15494$ & $<\vec{S}_{14} \cdot \vec{S}_{15}> = -0.61209$ \\
 & $<\vec{S}_2 \cdot \vec{S}_5> = -0.35762$ &  &  & \\
 & $<\vec{S}_5 \cdot \vec{S}_{11}> = -0.40890$ &  &  & \\
\end{tabular}
\label{table:11}
\end{center}
\end{table}

\begin{table}
\begin{center}
\caption{Lowest energies ($E_{0}$), multiplicities ($mult.$) and corresponding
irreducible representations ($irrep.$) in each total spin $S$ sector for the
clusters. The spatial irreducible representation notation follows Ref.
\cite{Altmann94}. Indices $s$ and $a$ indicate the behavior under spin
inversion, where $s$ stands for symmetric and $a$ for antisymmetric. It is only
possible to calculate them for states lying relatively low or high in the
energy spectrum. A comma is introduced when necessary to avoid confusion
between the notation for the spatial irreducible representation and the
behavior under spin inversion.}
\begin{tabular}{c|c|c|c|c|c|c|c|c|c}

 & $C_{24}$ & $C_{24}$ & $C_{24}$ & $C_{26}$ & $C_{26}$ & $C_{26}$ & $C_{28}$ & $C_{28}$ & $C_{28}$ \\
\hline
$S$ & $E_{0}$ & $mult.$ & $irrep.$ & $E_{0}$ & $mult.$ & $irrep.$ & $E_{0}$ & $mult.$ & $irrep.$ \\
\hline
0 & -11.71937 & 1 & $B_{1,s}$ & -12.60898 & 1 & $A_{2,a}^{'}$ & -13.57486 & 2 & $E_{s}$ \\
\hline
1 & -11.46814 & 2 & $E_{3,a}$ & -12.49297 & 2 & $E_{s}^{'}$ & -13.55978 & 3 & $T_{2,a}$ \\
\hline
2 & -10.92259 & 1 & $B_{1,s}$ & -12.13819 & 1 & $A_{2,a}^{'}$ & -13.42327 & 1 & $A_{1,s}$ \\
\hline
3 & -10.13390 & 1 & $B_{2,a}$ & -11.39120 & 1 & $A_{1,s}^{'}$ & -12.45085 & 3 & $T_{2,a}$ \\
\hline
4 & -8.88178 & 1 & $A_{1,s}$ & -10.16420 & 1 & $A_{2,a}^{'}$ & -11.45102 & 1 & $A_{1,s}$ \\
\hline
5 & -7.28948 & 2 & $E_{5}$ & -8.83042 & 1 & $A_{1,s}^{'}$ & -10.18889 & 1 & $A_{1,a}$ \\
\hline
6 & -5.49575 & 1 & $A_{1}$ & -7.01596 & 1 & $A_{2}^{'}$ & -8.51700 & 1 & $A_{1}$ \\
\hline
7 & -3.46016 & 1 & $B_{2}$ & -5.07125 & 1 & $A_{1}^{'}$ & -6.56848 & 3 & $T_{2}$ \\
\hline
8 & -1.24214 & 1 & $B_{1}$ & -2.91337 & 2 & $E^{'}$ & -4.52626 & 2 & $E$ \\
\hline
9 & 1.10955 & 1 & $A_{2}$ & -0.62134 & 2 & $E^{'}$ & -2.32108 & 1 & $A_{1}$ \\
\hline
10 & 3.59067 & 1 & $B_{1,s}$ & 1.75066 & 1 & $A_{2}^{'}$ & -0.049533 & 1 & $A_{1}$ \\
\hline
11 & 6.29289 & 2 & $E_{3,a}$ & 4.36198 & 1 & $A_{1}^{'}$ & 2.48354 & 3 & $T_{2}$ \\
\hline
12 & 9 & 1 & $A_{1,s}$ & 7.04289 & 1,2 & $A_{2,a}^{'},E^{'}$ & 5.11292 & 2 & $E$ \\
\hline
13 & & & & 9.75 & 1 & $A_{1,s}^{'}$ & 7.79289 & 1,3 & $A_{1,a},T_{2}$ \\
\hline
14 & & & & & & & 10.5 & 1 & $A_{1,s}$ \\
\hline
 & $C_{30}$ & $C_{30}$ & $C_{30}$ & $C_{32,I}$ & $C_{32,I}$ & $C_{32,I}$ & $C_{32,II}$ & $C_{32,II}$ & $C_{32,II}$ \\
\hline
$S$ & $E_{0}$ & $mult.$ & $irrep.$ & $E_{0}$ & $mult.$ & $irrep.$ & $E_{0}$ & $mult.$ & $irrep.$ \\
\hline
0 & -14.88742 & 1 & $A_{2,a}^{''}$ & -15.87092 & 1 & $A_{1u,s}$ & -15.93723 & 1 & $A_{1,s}^{''}$ \\
\hline
1 & -14.62495 & 1 & $A_{1,s}^{'}$ & -15.67299 & 2 & $E_{u,a}$ & -15.77366 & 1 & $A_{2,a}^{'}$ \\
\hline
2 & -14.16221 & 1 & $A_{2,a}^{'}$ & -15.38287 & 1 & $A_{1g,s}$ & -15.35876 & 1 & $A_{1}^{'}$ \\
\hline
3 & -13.47738 & 2 & $E_{2,s}^{'}$ & -14.72334 & 1 & $A_{2g}$ & -14.68665 & 1 & $A_{2}^{'}$ \\
\hline
4 & -12.52542 & 1 & $A_{2}^{''}$ & -13.83231 & 1 & $A_{1g}$ & -13.80765 & 1 & $A_{1}^{'}$ \\
\hline
5 & -11.26275 & 1 & $A_{1}^{''}$ & -12.66407 & 1 & $A_{2g}$ & -12.59064 & 1 & $A_{1}^{'}$ \\
\hline
6 & -9.71967 & 2 & $E_{2}^{''}$ & -11.25111 & 1 & $A_{1g}$ & -11.12132 & 1 & $A_{1}^{''}$ \\
\hline
7 & -7.94218 & 2 & $E_{2}^{'}$ & -9.56173 & 1 & $A_{2g}$ & -9.42233 & 1 & $A_{1}^{'}$ \\
\hline
8 & -6.07738 & 1 & $A_{2}^{''}$ & -7.65604 & 1 & $A_{1g}$ & -7.61188 & 1 & $A_{1}^{'}$ \\
\hline
9 & -3.98134 & 1 & $A_{1}^{''}$ & -5.59828 & 1 & $A_{2g}$ & -5.58108 & 1 & $A_{2}^{''}$ \\
\hline
10 & -1.69706 & 1 & $A_{2}^{'}$ & -3.40688 & 1 & $A_{1g}$ & -3.43626 & 1 & $A_{1}^{'}$ \\
\hline
11 & 0.66164 & 1 & $A_{1}^{'}$ & -1.06488 & 1 & $A_{2g}$ & -1.14953 & 1 & $A_{2}^{''}$ \\
\hline
12 & 3.08988 & 1 & $A_{2}^{''}$ & 1.37454 & 1 & $A_{2g}$ & 1.25805 & 1 & $A_{1}^{'}$ \\
\hline
13 & 5.68792 & 1 & $A_{1}^{'}$ & 3.84610 & 1 & $A_{2g}$ & 3.79378 & 1 & $A_{2}^{''}$ \\
\hline
14 & 8.42712 & 1 & $A_{2}^{''}$ & 6.44732 & 1 & $A_{1g}$ & 6.42049 & 1 & $A_{1}^{'}$ \\
\hline
15 & 11.25 & 1 & $A_{1,s}^{'}$ & 9.19098 & 1 & $A_{2g}$ & 9.19475 & 1 & $A_{2}^{''}$ \\
\hline
16 & & & & 12 & 1 & $A_{1g,s}$ & 12 & 1 & $A_{1,s}^{'}$ \\

\end{tabular}
\label{table:12}
\end{center}
\end{table}

%
%

\newpage

\end{document}